\documentclass[prl,showpacs,twocolumn]{revtex4}
\usepackage{mathrsfs}
\usepackage{amsfonts}
\usepackage{graphicx}
\usepackage{amsmath}
\usepackage{amssymb}
\begin{document}
\title{Shift of Dirac points and strain induced pseudo-magnetic field in graphene}
\author{Hua Tong Yang}\email{yanght653@nenu.edu.cn}
\affiliation{Center for Advanced Optoelectronic Functional Materials
Research, Key Laboratory for UV-Emitting Materials and Technology of
Ministry of Education, and School of Physics, Northeast Normal
University, Changchun 130024, China}

\begin{abstract}
We propose that the strain induced effective pseudo-magnetic field
in graphene can also be explained by a curl movement of the Dirac
points, if the Dirac points can be regarded as a slowly varying
function of position. We also prove that the Dirac points must be
confined within two triangles, each one has 1/8 the area of the
Brillouin zone.
\end{abstract}
\pacs{73.22.Pr,  73.22.Dj, 73.22.Gk, 73.20.At}

\maketitle

The discovery of graphene, a monolayer carbon atom sheet
\cite{Novoselov-sci306}, and the development of experimental
technique to manipulate this two-dimensional(2D) material have
ignited intense interest in this system
\cite{Geim,Meyer,Vozmediano,Neto}. One of the most attractive
characters of graphene is that its low energy excitation satisfies a
massless 2D Dirac equation \cite{Semenoff}, and the chemical
potential crosses its Dirac points(or Fermi points) in neutral
graphene. These special characters lead to many unusual properties
and new phenomena \cite{Gusynin,Neto,Novoselov2005, Zhang2005}, such
as the anomalous integer quantum Hall effect(QHE)
\cite{Novoselov2005, Zhang2005}. Recently, experiments have
confirmed another remarkable effect that mechanical strain can
induce a very strong effective pseudo-magnetic field, leading to a
pseudo-QHE, which can be observed in zero magnetic field
\cite{Guinea-10,Levy}. In this paper we propose that the strain
induced effective vector potential can also be explained by shift
$\delta \mathbf{K}(\mathbf{x})$ of the Dirac points
$\mathbf{K}(\mathbf{x})$, its effective pseudo-magnetic field is in
proportion to $\nabla\times\mathbf{K}(\mathbf{x})$, only if the
Dirac points $\mathbf{K}(\mathbf{x})$ can be regarded as a slowly
varying function of position, and the Fermi velocity is generalized
to a tensor\cite{Zhu-Wang}. We also prove that the Dirac points can
not be arbitrarily moved, they must be confined within two
triangles, each one has 1/8 the area of the Brillouin zone(BZ).

\begin{figure}[htb]
\setlength{\unitlength}{1cm}
\begin{center}\small
\begin{picture}(8,3.7)
\includegraphics*[width=8cm]{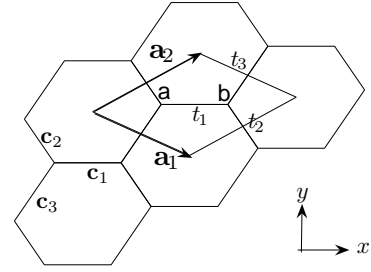}
\end{picture}
\end{center}
\caption{\label{cell}Unit cell and hopping parameters for deformed graphene.} 
\end{figure}

Firstly, consider a tight-binding Hamiltonian describing a uniformly
deformed honeycomb lattice with three different nearest-neighbor
hopping energies $t_1,t_2,t_3$\cite{Hasegawa, Montambaux08,
Montambaux09}:
\begin{eqnarray} \label{Hamilton}
\hat{H}=-\sum_{<\mathbf{i}a,\mathbf{j}b>}t_{\mathbf{i}a,\mathbf{j}b}
c^{\dag}_{\mathbf{i}a}c_{\mathbf{j}b}+h.c.,
\end{eqnarray}
where $c_{\mathbf{j}b}$ ($c^{\dag}_{\mathbf{i}a}$) are
annihilation(creation) operators, $\mathbf{i}$($\mathbf{j}$) are
position vectors of unit cells, $a$($b$) denote two inequivalent
atoms in a unit cell, $t_{\mathbf{i}a,\mathbf{j}b}$ is the
electronic hopping energy from the $\mathbf{j}$th unit cell $b$ atom
to $\mathbf{i}$th unit cell $a$ atom. Suppose that the deformed
lattice remains invariant under spatial translation, i.e.,
$t_{\mathbf{i}a,\mathbf{j}b}$ only depends on ${\bf i}-{\bf j}$, but
the three nearest-neighbor hopping energies $t_{1,2,3}$ may be
different owing to anisotropy of strains, as shown in
Fig.\ref{cell}.
The hopping parameters can be written as some $2\times2$ matrixes
$\mathbf{t}(\mathbf{i}-\mathbf{j})$, whose elements are defined by
$[{\bf t}(\mathbf{i}-\mathbf{j})]_{a,b}\equiv
t_{\mathbf{i}a,\mathbf{j}b}.$ For this nearest-neighbor
tight-binding Hamiltonian, the non-vanishing hopping matrixes are
\begin{eqnarray}{\bf t}(0)=\bigg(\begin{array}{cc}0&t_1\\
t_1&0\end{array}\bigg),{\bf t}({\bf a}_1)=\bigg(\begin{array}{cc}0&t_2\\
0&0\end{array}\bigg),{\bf t}({\bf a}_2)=\bigg(\begin{array}{cc}0&t_3\\
0&0\end{array}\bigg)\end{eqnarray} and ${\bf t}(-{\bf a}_1)={\bf
t}^\dag({\bf a}_1),{\bf t}(-{\bf a}_2)={\bf t}^\dag({\bf a}_2).$ By
Fourier transformation
$$c_{\mathbf{j},a(b)}=\frac{1}{\sqrt{N}}\sum_{\mathbf{k}}
c_{\mathbf{k},a(b)}\exp(i\mathbf{k}\cdot\mathbf{j})$$ with $N$ a
normalization constant, the Hamiltonian (\ref{Hamilton}) can be cast
into the form
\begin{eqnarray}\label{Hk}
\hat{H}=-\sum_{\mathbf{k}}\big[c^{\dag}_{\mathbf{k},a},c^{\dag}_{\mathbf{k},b}\big]
\bigg[\begin{array}{cc}h_{aa}(\mathbf{k})&h_{ab}(\mathbf{k})\\
h_{ba}(\mathbf{k})& h_{bb}(\mathbf{k})\end{array}\bigg]
\bigg[\begin{array}{c}c_{\mathbf{k},a}\\c_{\mathbf{k},b}\end{array}\bigg],
\end{eqnarray}
where $h_{aa}(\mathbf{k})=h_{bb}(\mathbf{k})=0,$
$h_{ab}(\mathbf{k})=h_{ba}^{\ast}(\mathbf{k}),$ and
\begin{eqnarray}h_{ba}(\mathbf{k})=t_1+t_2\exp(i\mathbf{k}\cdot\mathbf{a}_1)
+t_3\exp(i\mathbf{k}\cdot\mathbf{a}_2),\end{eqnarray} with
$\mathbf{a}_1,$$\mathbf{a}_2$ the lattice unit vectors. The energy
bands obtained by diagonalizing this Hamiltonian are\cite{Wallace}
\begin{equation}\label{energyband}E_{\pm}(\mathbf{k})=\pm|t_1+\tilde{t}_2(\mathbf{k})+\tilde{t}_3(\mathbf{k})|,\end{equation}
where
$\tilde{t}_{2}(\mathbf{k})=t_{2}e^{i\mathbf{k}\cdot\mathbf{a}_1},$
$\tilde{t}_{3}(\mathbf{k})=t_{3}e^{i\mathbf{k}\cdot\mathbf{a}_2}$,
the plus sign corresponds to the upper($\pi$) and minus to the
lower($\pi^*$) band respectively. From Eq.(\ref{energyband}) we
notice that if $\mathbf{K}$ is a zero point of $h_{ba}(\mathbf{K})$,
i.e.,
\begin{equation}\label{t123}t_1+\tilde{t}_2(\mathbf{K})+\tilde{t}_3(\mathbf{K})=0,\end{equation}
then $E_+(\mathbf{k})$ and $E_{-}(\mathbf{k})$ will meet at
$\mathbf{K}$, i.e., $E_{+}(\mathbf{K})=E_{-}(\mathbf{K})=0$, this
$\mathbf{K}$ is known as the Dirac point. The Hamiltonian (\ref{Hk})
can be expanded up to a linear order in
$\mathbf{p}=\mathbf{k}-\mathbf{K}$ in a neighborhood of point
$\mathbf{K}$
\begin{eqnarray}\label{Linear H}
\bigg[\begin{array}{cc}h_{aa}(\mathbf{k})&h_{ab}(\mathbf{k})\\
h_{ba}(\mathbf{k})&h_{bb}(\mathbf{k})\end{array}\bigg]\simeq\bigg[\begin{array}{cc}0&\vec{\alpha}^*\cdot\mathbf{p}\\
\vec{\alpha}\cdot\mathbf{p}&0\end{array}\bigg]=v_{\mu\nu}\sigma^{\mu}p^{\nu},
\end{eqnarray}
where $\mu,\nu=1,2$ denote two components of a 2D vector and a sum
over the repeated indices $\mu,\nu$ is implied, $\vec{\alpha}$ is a
complex vector with $\textrm{Re}(\vec{\alpha})=(v_{11},v_{12})$,
$\textrm{Im}(\vec{\alpha})=(v_{21},v_{22})$, $\sigma^{1,2}$ are
Pauli matrixes acting on the sublattice degree of freedom, tensor
$v_{\mu\nu}$ represents the anisotropy of the dispersion near the
Dirac points, it only occurs noticeable departure from
$v_F\delta_{\mu\nu}$ in a strongly deformed graphene\cite{Zhu-Wang}.
However, after this modification the strain induced effective vector
potential will acquire a direct physical meaning. For a graphene
under nonuniform but slowly varying strain, $t_i(\mathbf{x})$ and
hence the Dirac point $\mathbf{K}(\mathbf{x})$ as well as
$v_{\mu\nu}(\mathbf{x})$ can be regarded as some smooth functions of
position $\mathbf{x}$, the local linearized Hamiltonian
$v_{\mu\nu}\sigma^{\mu}(k^{\nu}-K^{\nu}(\mathbf{x}))$ on the RHS of
Eq.(\ref{Linear H}) can be cast into
\begin{eqnarray}\label{H_eff}v_{\mu\nu}(\mathbf{x})\sigma^{\mu}(p^{\nu}-\delta
K^{\nu}(\mathbf{x})),\end{eqnarray} where
$\mathbf{p}-\delta\mathbf{K}(\mathbf{x})\equiv\mathbf{k}-\mathbf{K}(\mathbf{x})$,
$\delta
\mathbf{K}(\mathbf{x})\equiv\mathbf{K}(\mathbf{x})-\mathbf{K}_f$
with $\mathbf{K}_f$ the corresponding Dirac point in strain-free
graphene. Unlike the usual explanation of the strain induced gauge
field in graphene\cite{Kane, Neto}, where the effective vector
potential is an auxiliary quantity and describes the mixed effects
of both anisotropy of $v_{\mu\nu}$ and the shift of Dirac point,
here the vector potential only represents the relative translation
of the Dirac points, $(e/c)\mathbf{A}(\mathbf{x})=\delta
\mathbf{K}(\mathbf{x})$, its pseudo-magnetic field
$\mathbf{B}(\mathbf{x})=(c/e)\nabla\times \mathbf{K}(\mathbf{x})$,
and the physical effects are mainly determined by the
pseudo-magnetic flux through a loop  $(c/e)\oint_L
\mathbf{K}(\mathbf{x})\cdot d\mathbf{x}$. In the following sections
we shall discuss the properties of $\mathbf{K}(\mathbf{x})$, and
illustrate how a curl field $\mathbf{K}(\mathbf{x})$ is induced by a
strain.

\begin{figure}[htb]
\begin{center}\small
\setlength{\unitlength}{1cm}
\begin{picture}(8,5)
\includegraphics*[width=8cm]{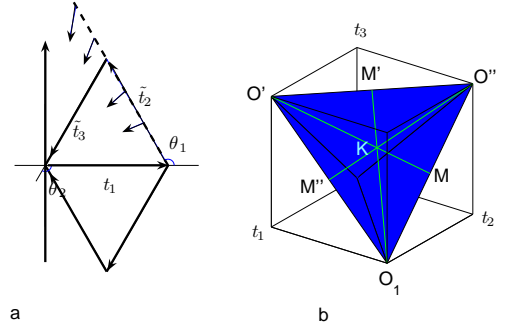}
\end{picture}
\end{center}
\caption{\label{domain}(color online). (a) Zero points of
$h_{ba}(\mathbf{k})$ determine two directed triangles with edges
$t_{1,2,3}$ in the complex plane. For a given $t_1$ and a fixed
direction of $\tilde{t}_2$, the arguments of $\tilde{t}_3$ must
satisfy conditions (\ref{range}) to ensure $t_{2,3}\geq 0$. (b)
Dirac points exist if $t_i$ satisfy inequalities (\ref{condition}),
which describe a pyramidal domain in $(t_1,t_2,t_3)$ space, if
$(t_1,t_2,t_3)$ goes beyond this domain, an energy gap will be
opened.}
\end{figure}

\begin{figure}[hbt]
\begin{center}\small
\setlength{\unitlength}{1cm}
\begin{picture}(8,6.7)
\includegraphics*[width=8cm]{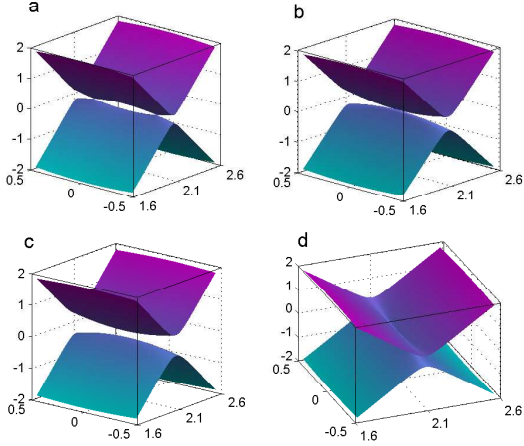}
\end{picture}
\end{center}
\caption{\label{bandgap}(color online) (a) Energy band when two
Dirac points are very close, where $t_1=2.8, t_{2,3}=1.45,$ (b), (d)
$t_1=2.8, t_{2,3}=1.4,$ two Dirac points are equivalent(superposed),
(c) $t_1=2.8, t_{2,3}=1.35,$ an energy gap occurs.}
\end{figure}

From Eq.(\ref{t123}) we know that the vectors representing $t_1$,
$\tilde{t}_2(\mathbf{K})$, $\tilde{t}_3(\mathbf{K})$ in the complex
plane can form a directed triangle for a Dirac point $\mathbf{K}$,
as illustrated in Fig.\ref{domain}a.
According to the triangle inequality, we have the following
necessary and sufficient conditions for the existence of the Dirac
points\cite{Hasegawa}:
\begin{equation}\label{condition}t_1+t_2\geq t_3,~t_2+t_3\geq t_1,~t_3+t_1\geq t_2.\end{equation}
These conditions define a pyramidal domain in the $(t_1,t_2,t_3)$
space, shown in Fig.\ref{domain}b. If $t_1,t_2,t_3$ satisfy
inequalities (\ref{condition}), then there exists two directed
triangles with the same edges $t_1,t_2,t_3$ but different possible
orientations, which determine two angles $\theta_1, \theta_2$
satisfying $t_1+ t_2e^{i\theta_1}+t_3e^{i\theta_2}=0$, where
$\theta_1, \theta_2$ are given by the law of cosine
\begin{eqnarray}\label{t-K map}\begin{split}
\theta_{\pm1}&=\pm\big[\pi-\arccos\big(\frac{t_1^2+t_2^2-t_3^2}{2t_1
t_2}\big)\big],\\
\theta_{\pm2}&=\pm\big[\arccos\big(\frac{t_1^2+t_3^2-t_2^2}{2t_1
t_3}\big)-\pi\big].\end{split}
\end{eqnarray}
Thus the Dirac points $\mathbf{K}$ can be determined by letting
\begin{equation}\exp(i\mathbf{K}\cdot \mathbf{a}_1)=\exp(i\theta_1), ~\exp(i\mathbf{K}\cdot \mathbf{a}_2)=\exp(i\theta_2),\end{equation}
so we have
\begin{equation}\label{K-components}
\mathbf{K}=\frac{1}{2\pi}\big(\theta_1\mathbf{b}_1+\theta_2\mathbf{b}_2\big)+\mathbf{K}_{0},
\end{equation}
with $\mathbf{b}_1,\mathbf{b}_2$ the reciprocal lattice vectors
defined by $\mathbf{a}_i\cdot\mathbf{b}_j=2\pi \delta_{ij}$, and
$\mathbf{K}_{0}=n\mathbf{b}_1+m\mathbf{b}_2$ with $n,m$ are
arbitrary integers. Notice that if $t_1+\tilde{t}_2+\tilde{t}_3=0$,
then $t_1+\tilde{t}^*_2+\tilde{t}^*_3=0$, this implies that there
exists two Dirac points $\mathbf{K}(\mathbf{x})$ and
$-\mathbf{K}(\mathbf{x})$. However, if $(t_1,t_2,t_3)$ exactly
locates on the boundary surface of the pyramid, e.g., $t_1=t_2+t_3$,
then the two triangles will mutually coincide and
$\tilde{t}_2=\tilde{t}^*_2$, $\tilde{t}_3=\tilde{t}^*_3$(see
Fig.\ref{domain}a), hence $\mathbf{K}(\mathbf{x})$ and
$-\mathbf{K}(\mathbf{x})$ become equivalent, and
$\vec{\alpha}=i(t_2\mathbf{a}_1+t_3\mathbf{a}_2)$ becomes a pure
imaginary vector, so the Fermi velocity in the directions
perpendicular to $\vec{\alpha}$ vanishes(Fig.\ref{bandgap}b and
\ref{bandgap}d)\cite{Montambaux08, Montambaux09,Pereira}. If
$(t_1,t_2,t_3)$ goes beyond the domain defined by
Eq.(\ref{condition}), e.g., $t_1>t_2+t_3$, Eq.(\ref{t123}) will have
no any root, an energy gap with magnitude $E_{g}=2(t_1-t_2-t_3)$
will occur at the corresponding points
$\mathbf{K}_{\pm}=\pm1/2(\mathbf{b}_1+\mathbf{b}_2)$(see
Fig.\ref{bandgap}c)\cite{Zhou}, and the effective Hamiltonian
(\ref{H_eff}) must be further modified by adding a mass term and
some second order terms.

\begin{figure}[tbh]
\setlength{\unitlength}{1cm}
\begin{center}\begin{picture}(8,7)\small
\includegraphics*[width=8cm]{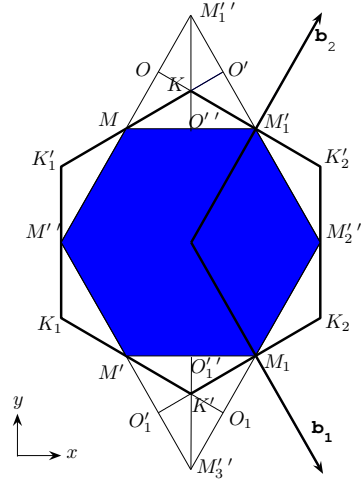}
\end{picture}\end{center}
\caption{\label{Brillouin}(color inline) Rang of the Dirac points
consists of six triangles in first the BZ, or $\bigtriangleup
MM'_1M''_1$ and $\bigtriangleup M'M_1M''_3$.}
\end{figure}

Another important property is the range of $\mathbf{K}(\mathbf{x})$.
We shall prove that the Dirac points must be confined within some
special regions of the BZ. To this end, notice that if a Dirac point
$\mathbf{K}=(1/2\pi)(\theta_1,\theta_2)$ is given, then its
associated $t_{1,2,3}$ can also be determined up to an arbitrary
factor, except six special cases of $\theta_1, \theta_2 =0,
\pm\pi$(see Fig. \ref{domain}a). If $\theta_1,\theta_2\neq 0,
\pm\pi$. According to the law of sines we have
\begin{eqnarray}\label{K-t map}
\frac{t_2}{t_1}=\frac{\sin \theta_2}{\sin(\theta_1-\theta_2)},~~
\frac{t_3}{t_1}=\frac{\sin\theta_1}{\sin(\theta_2-\theta_1)},
\end{eqnarray}
or \begin{eqnarray}\label{K-t map1}(t_1,t_2,t_3)\propto
(\sin(\theta_2-\theta_1),-\sin\theta_2,\sin\theta_1).\end{eqnarray}
For the six special cases we have: if $(\theta_1,
\theta_2)=\pm(\pi,\pi),$ $t_1=t_2+t_3;$ if $(\theta_1,
\theta_2)=\pm(\pi,0),$ $t_2=t_1+t_3;$ if $(\theta_1,
\theta_2)=\pm(0,\pi),$ $t_3=t_1+t_2$. From Eq.(\ref{K-t map1}) we
can find that the $\theta_1,\theta_2$ must satisfy some constrain
conditions to guarantee $t_{1,2,3}\geq 0$, as illustrated in
Fig.\ref{domain}a. For an arbitrary $t_2$ and a fixed
$\theta_1$(direction of $\tilde{t}_2$), $\tilde{t}_3$ must point in
a direction between the directions of $-\tilde{t}_2$ and negative
real axis, i.e., argument $\theta_1, \theta_2$ must satisfy
\begin{eqnarray}\label{range}\begin{split}
\theta_1&+\pi<\theta_2<\pi, &\theta_1&\in(-\pi,0),\\
-\pi&<\theta_2<\theta_1-\pi,&\theta_1&\in(0,\pi).\\
\end{split}\end{eqnarray}
These two inequalities respectively determine the range of
$\mathbf{K}(\mathbf{x})$ and $-\mathbf{K}(\mathbf{x})$. They
describe two open triangles $\bigtriangleup MM'_1M''_1$ and
$\bigtriangleup M'M_1M''_3$  in reciprocal space, as shown in
Fig.\ref{Brillouin}, each one has 1/8 the area of a unit cell of the
reciprocal space(the parallelogram $M''M''_1M''_2M''_3$), and each
Dirac point is confined within a triangle, so, the Dirac points
$\mathbf{K}$ and $-\mathbf{K}$ can meet(become equivalent) only at
the vertexes of $\bigtriangleup MM'_1M''_1$ and $\bigtriangleup
M'M_1M''_3$. The remaining hexagon(blue in Fig.\ref{Brillouin}) is a
forbidden region for the Dirac points. This confinement also limits
the order of magnitude of $\nabla\times\mathbf{K}(\mathbf{x})$,
i.e., the strain induced pseudo-magnetic field.
In order to show the underlying regularity, here we have ignored the
variations of $\mathbf{b}_1,\mathbf{b}_2$ with the deformation of
lattice, and simply sketch all $\mathbf{K}=(k_1,k_2)$ in the same
affine frame. After translating to the first BZ of graphene,
$\bigtriangleup MM'_1M''_1$ and $\bigtriangleup M'M_1M''_3$ are
equivalent to a ringlike region consists of six triangles $\triangle
MKM'_1,$ $\triangle M_1K_2M''_2,$ $\triangle M''K_1M',$ etc.

\begin{figure}[htb]
\begin{center}
\small\setlength{\unitlength}{1cm}
\begin{picture}(8.5,6.6)
\includegraphics*[width=8.5cm]{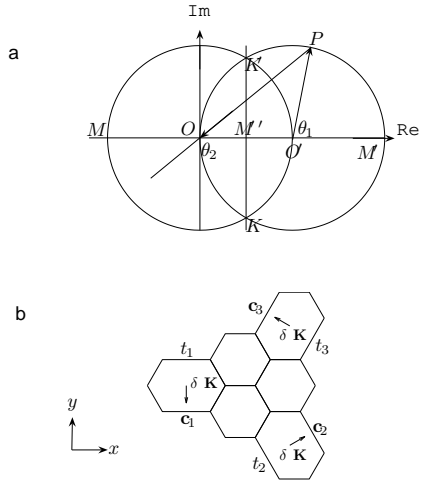}
\end{picture}
\end{center}
\caption{\label{shift}(a) The trajectories of P in three ideal
cases. (b) A schematic diagram of the shift of Dirac point
$\delta\mathbf{K}$ and its curl, $\delta\mathbf{K}$ perpendicular to
$\mathbf{c}_i$, if bond $\mathbf{c}_i$ is slightly elongated.}
\end{figure}

In order to illustrate how a non-vanishing $\nabla\times\mathbf{K}$
is induced by strain, we only need to analyze three ideal cases, in
which only one $t_i$ is slightly changed, $t_i\rightarrow t_0+\delta
t_i$, while the other two $t_{j,k}$ remain constant, $t_j=t_k=t_0$,
which can also be roughly regarded as that the bond $\mathbf{c}_i$
is elongated(or compressed) while the other two bonds
$\mathbf{c}_j,\mathbf{c}_k$ and their directions remain fixed(see
Fig.\ref{shift}b). Notice that the Dirac points only depend on the
relative proportions of $t_1,t_2,t_3$, so, as an equivalent case, we
can always assume that $t_1$ remains constant and only $t_2$, $t_3$
are variables. Moreover, in these equivalent cases the $\tilde{t}_2$
and $\tilde{t}_3$ can be determined by the end of the vector
$t_1+\tilde{t}_2=-\tilde{t}_3$, denoted by $P$ in Fig.\ref{shift}a.
So, we can represent the variation of the Dirac points by the shift
of the point P. To this end, we have to determine the corresponding
P of the three classes of characteristic points in the range of the
Dirac points: (1) $K$(or $K'$) etc.(see Fig.\ref{Brillouin}),
according to Eq.(\ref{K-t map}), Dirac points locate at these two
points only if $t_1=t_2=t_3$, their corresponding P is located at
$K$(or $K'$) in Fig.\ref{shift}a; (2) critical points $M(M_1)=(\pm
1/2,0)$, $M'(M'_1)=(0,\pm 1/2)$, and $M''(M''_2)= (\pm 1/2, \pm
1/2),$ in these cases there exist only one Dirac point since the
points $\mathbf{K}$ and $-\mathbf{K}$ are equivalent, their
corresponding $(t_1,t_2,t_3)$ are located on the boundary of the
pyramidal domain, while their corresponding P are located at the
real axis in Fig.\ref{shift}a; (3) $O$, $O'$, $O''$ etc., their
corresponding P are the centers of two circles and the infinite
limit points of the straight line $KK'$ in Fig.\ref{shift}a, which
respectively correspond to the limits of $t_2\rightarrow 0$,
$t_3\rightarrow 0$ and $t_1\rightarrow 0$(equivalent to
$t_2=t_3\rightarrow\infty$).
 Now we analyze the shifts of the Dirac points in the three ideal situations. (1)
$t_1,t_2$ remain constant and $t_2=t_1$, only $t_3$ is variable, the
trajectory of the corresponding P is a circle with radius $t_1$ and
centered at the point $(t_1,0)$ in Fig.\ref{shift}a, so the
arguments $(\theta_1,\theta_2)$ satisfy
\begin{eqnarray}\label{t1=t2}\begin{split}
\theta_1-2\theta_2-2\pi=0,&~~\theta_1\in (-\pi,0),\\
\theta_1-2\theta_2+2\pi=0,&~~\theta_1\in
(0,\pi).\end{split}\end{eqnarray} They describe line segments
$M'O_1$ and $M'_1O$ in Fig.\ref{Brillouin}; (2) $t_3(=t_1)$ remain
constant while $t_2$ is variable, the trajectory of corresponding P
is another circle with radius $t_1$ centered at the origin, its
associated $(\theta_1,\theta_2)$ satisfy
\begin{eqnarray}\label{t1=t3}\begin{split}
\theta_2-2\theta_1+2\pi=0,&~~\theta_2 \in(-\pi,0),\\
\theta_2-2\theta_1-2\pi=0,&~~\theta_2\in
(0,\pi),\end{split}\end{eqnarray} which describe $MO'$ and $M_1O'_1$
in Fig.\ref{Brillouin}; (3) $t_1$ remains constant while $t_2,t_3$
are variable but $t_2=t_3$(or vice versa, $t_1$ is variable,
$t_2(=t_3)$ remain constant), the trajectory of P is straight line
$KK'$, $(\theta_1,\theta_2)$ satisfy
\begin{equation}\label{t2=t3}\theta_1+\theta_2=0, ~~\frac{\pi}{2}<|\theta_1|<\pi,\end{equation}
which describe $M''_1O''$ and $M''_3O''_1$ in Fig.\ref{Brillouin}.
Summarizing Eqs.(\ref{t1=t2})(\ref{t1=t3})(\ref{t2=t3}) and
comparing with Fig.\ref{Brillouin}, we observe that if a band, e.g.,
$\mathbf{c}_1$ is slightly elongated (or compressed) along its
direction, $\mathbf{c}_1\rightarrow (1+\delta)\mathbf{c}_1$, while
the other two bonds $\mathbf{c}_2, \mathbf{c}_3$ remain fixed, then
$t_1$ will be slightly changed while $t_2(=t_3)$ remain constant,
the Dirac point $\mathbf{K}$ will be slightly moved in the direction
perpendicular to $\mathbf{c}_1$, i.e., $\delta K_y\neq0$(see
Fig.\ref{Brillouin}, $K$ moves towards $O''$ if $t_1$ decreases,
towards $M''_1$ if $t_1$ increases), $\mathbf{K'}(=-\mathbf{K})$ is
moved in the opposite direction. Thus, if the elongation of
$\mathbf{c}_1$ is slowly varying in the x-direction, i.e, $\partial
t_1/\partial x\neq 0$, then $\partial K_y/\partial x\neq 0$, the
other two cases are similar. So, a nonuniform strain as
schematically shown in Fig.\ref{shift}b can induce a curl field
$\mathbf{K}(\mathbf{x})$, $\nabla\times \mathbf{K}\neq 0$.

\begin{acknowledgments} We thank Yugui Yao, Chengshi Liu, Yichun
Liu for their helpful discussions. This work was supported by the
National Science Foundation of China(Grant Nos.10974027, 50725205,
50832001).
\end{acknowledgments}

\end{document}